# Ligand Field Exciton Annihilation in Bulk $CrCl_3$

Samanvitha Sridhar,[1,2] Ario Khansari,[1,2] Shaun, O'Donnell,[3] Alexandra T. Barth, [1,3] Evgeny O. Danilov,[1,3] Felix N. Castellano,[1,3] Paul A. Maggard,[3] and Daniel B. Dougherty[1,2]

1. Organic and Carbon Electronics Laboratories, North Carolina State University, Raleigh, NC 27695
2.Department of Physics, North Carolina State University, Raleigh, NC 27695
3.Department of Chemistry North Carolina State University, Raleigh, NC 27695-8204

The layered van der Waals material $CrCl_3$ exhibits very strongly bound ligand field excitons that control optoelectronic applications and are connected with magnetic ordering by virtue of their *d*-orbital origin. Time-resolved photoluminescence of these exciton populations at room temperature shows that their relaxation is dominated by exciton-exciton annihilation and that the spontaneous decay lifetime is very long. These observations allow the rough quantification of the exciton annihilation rate constant and contextualization in light of a recent theory of universal scaling behavior of the annihilation process.

## 1. Introduction

Chromium trihalides of the form $CrX_3$ have recently brought magnetic functionality into the world of 2D materials.[1-3] Single-layer ferromagnetic ordering has been observed in $CrI_3$, and thickness-based control of magnetic ordering has been demonstrated.[2] Furthermore, magnetism in this material can be controlled with an applied electric field[4] and both $CrI_3$[5] and $CrCl_3$[6] can incorporated into van der Waals heterostructures for spintronic applications. These transition metal halide-based materials exist within a family of related layered van der Waals compounds[7] including the possible spin liquid $RuCl_3$[8,9] and its Ir[10,11] and Cr[12,13] substituted variants.

The optoelectronic properties of the chromium trihalides add functionality to the layered magnetic van der Waals materials class.[14] Interest in the optical properties of chromium-based materials has a long history[15] connected to the canonical example of ruby phosphorescence.[16] Emissive properties associated with $Cr^{3+}$ can often be tuned by chemical composition and structure to select for energy and tune between broad fluorescence and narrow phosphorescence.[17] In the layered trihalides, optical absorption can be tuned across the visible spectrum by mixed halide composition[18] and the materials exhibit substantial photoconductivity[14] connected to magnetic ordering.[19] The intriguing fact that optical properties are intertwined with magnetic ordering is a major source of interest in the $CrX_3$ class. This is exemplified by the spontaneous circular polarization in photoluminescence from single layer $CrI_3$ in its ferromagnetic phase.[20] The connection between magnetism and optical properties arises due to the *d-d* origin of low energy optical excitations in chromium compounds, sometimes referred to as ligand field excitons (LFE's).

Ligand field excitons can be conceptualized using the ligand field splitting diagram in Figure 1a, where the qualitative effect of the presence of halide ligands is to break the 5-fold degeneracy of the $Cr^{3+}$ ion *d* orbitals into $t_{2g}$ and $e_g$ orbital sets. The lowest energy optical excitations involve exciting one of the 3d electrons across the ligand field gap from a $t_{2g}$ to an $e_g$ level. The result is an excited electron in an $e_g$ orbital and a hole left behind in the $t_{2g}$ orbital. Together, they comprise a quasiparticle that is spatially located primarily on a single $Cr^{3+}$ ion site called a LFE. Sophisticated many-body calculations[21] support the existence of this very small LFE with a strong exciton binding energy arising from the close proximity of the electron and hole.

The LFE's in $Cr^{3+}$ materials are textbook case studies in the optical properties associated with *d-d* excitations in transition metal compounds and often serve as an introduction to the use of Tanabe-Sugano diagrams.[22] The electronic structure of octahedral $Cr^{3+}$ is dictated by the local site geometry, which influences the crystal field strength *Dq* and the electrostatic interaction parameters described by the Racah parameters *B* and *C*. Coordination compounds involving this ion typically show a characteristic two-transition structure in the visible region of the steady state optical absorption spectra that arises from the two lowest energy spin-allowed excitations across the ligand field gap. When simplified to octahedral symmetry in point group *O*, these transitions are assigned to a lower energy transition of $^4A_2 \rightarrow {}^4T_2(F)$ and a higher energy transition of $^4A_2 \rightarrow {}^4T_1(F)$. The two LFE excitations differ from one another due to the differing coulomb repulsions between the excited electron in the $e_g$ orbital and the remaining two electrons in the $t_{2g}$ orbitals.[23] The assignment of these excitations is well understood,[15] but less attention has been paid to the mechanism of excited state relaxation time scales in chromium trihalide solids, especially in the context of the quasiparticle view of LFE's.

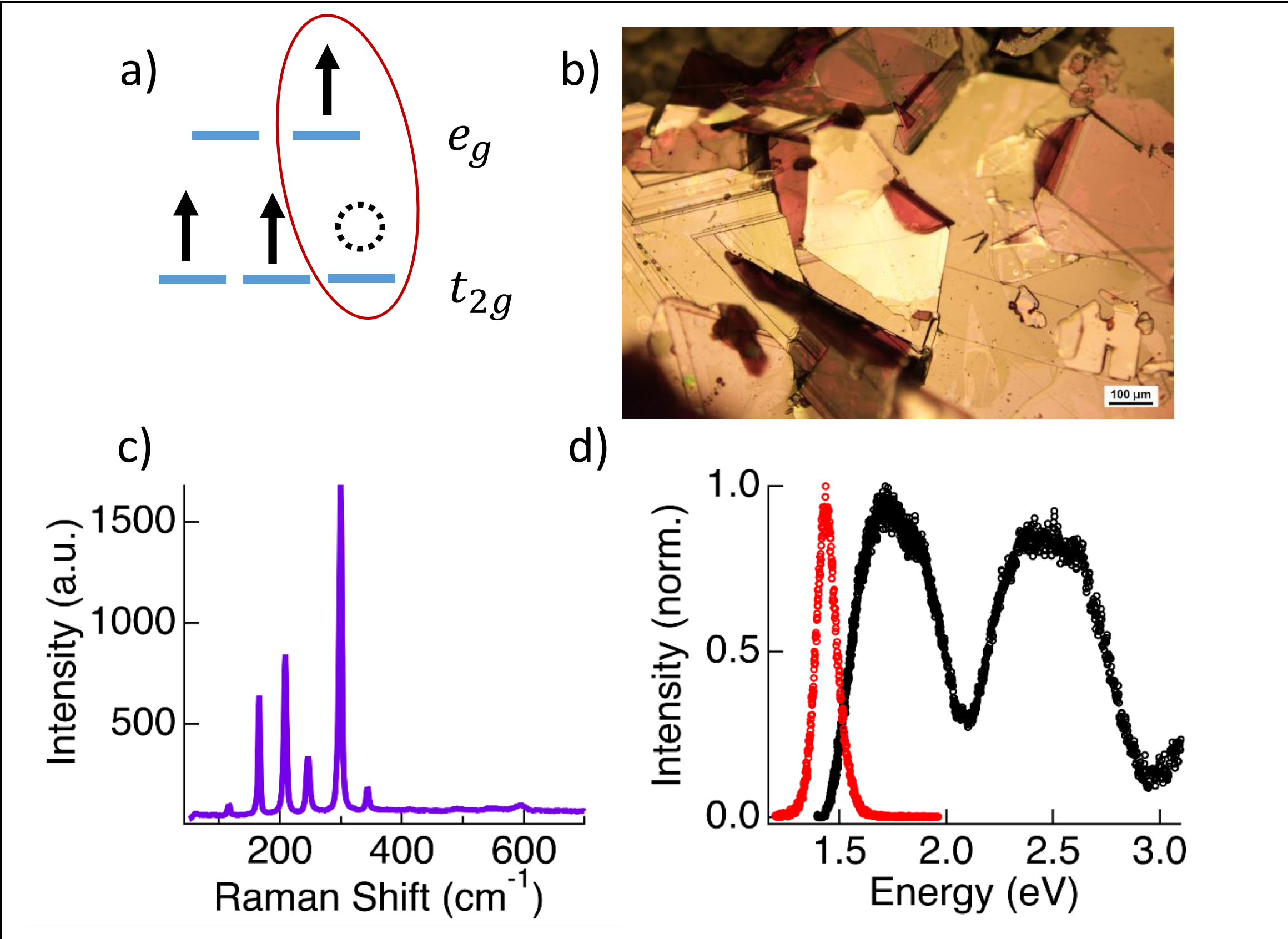

**Figure 1.** a) Schematic depiction of a Ligand Field Exciton formed by excitation of a ground state electron in a $t_{2g}$ orbital into an $e_g$ orbital, leaving a hole behind in the $t_{2g}$; b) Microscope image of the surface of one of the $CrCl_3$ crystals used in this work; c) Raman spectrum measured on one of the $CrCl_3$ crystals; d) Combined photoluminescence (532 excitation) and absorption spectra for $CrCl_3$.

The detailed behavior of LFE's is essential not only for its connection to the magneto-optics of 2D magnetic materials but also for contributing to the rich "landscape" of excitonic physics in strongly correlated materials more broadly.[24] This landscape includes LFE's and Mott-Hubbard excitons,[25-27] charge transfer excitons,[24] and even more exotic spin-orbit excitons.[28] This diverse set of relatively localized excited states is an important window into the possible phases, both steady and transient, of strongly correlated matter. For example, there has been an ongoing investigation into the possibility that excitonic effects contribute to the pairing mechanism in high-temperature superconductivity.[29,30] Recent inelastic X-ray scattering experiments explore this possibility by quantifying how *d-d* excitons in copper oxide materials interact with other degrees of freedom in the solid.[31] From this perspective, the time-resolved relaxation of *d-d* exciton in correlated materials could be an essential puzzle piece in explaining high-temperature superconductivity and seeking new many-body effects in other materials like the transition metal halides we focus on here.

Generally, LFEs can be studied by optical methods analyzing their absorption and emission spectra. Time-resolved spectroscopy adds another dimension to that analysis. The ligand field strength determines the emission mechanism of $Cr^{3+}$, quantified in terms of $Dq$ and $B$. The $Cr^{3+}$ excited state energies are significantly influenced by site symmetry and metal-ligand bond lengths, modulating the excited state ordering through changes in the local coordination environment. Recently, time-resolved photoluminescence spectroscopy of $CrCl_3$ (as well as other trihalides and their lanthanide-doped derivatives) has been measured at cryogenic temperatures.[32] Long first-order lifetimes of ~ 10 μs were observed under these conditions, indicating the possibility for excited state manipulation.

In this paper, we quantify the relaxation of LFE populations at room temperature in bulk $CrCl_3$ crystals by time-resolved photoluminescence (tr-PL). We find that excited state populations persist for time scales of several microseconds. Detailed quantitative analysis of the population decay shows pure second-order kinetics and demonstrates that the decay mechanism is exciton-exciton annihilation (EEA) of interacting LFEs. The significance of the finding is that LFEs at low density can persist for extended timescales, which may allow them to be manipulated in optoelectronic devices. However, the achievable exciton density is limited, and it may be necessary to develop tactics to mitigate annihilation.

## 2.Experimental Methods

Anhydrous $CrCl_3$ (Millipore Sigma) was loaded as received as 1-2 mm platelets into quartz ampoules, evacuated, and sealed under vacuum. The sealed ampoules were heated to 600-630 °C in a furnace for several days to recrystallize and enlarge the flakes. In some cases, recrystallization was repeated several times to increase the number of larger flakes with dimensions up to 2-3 mm. Crystalline samples were extracted from the ampoules, and, for most measurements, several 1-3 mm flakes were arranged randomly on double-sided carbon tape on a soda glass microscope slide.

Diffuse reflectance spectra (DRS) were measured at room temperature using a commercial assembly (Ocean Optics DR probe) that includes a tungsten-halogen light source with a fiber probe placed at a 45 degree angle relative to the incident excitation direction and a fiber optic cable couples the reflected light to a spectrometer (Avantes ULS2048L). The measured reflectance is converted to a quantity approximately proportional to the absorption coefficient by applying the Kubelka-Munk (K-M) method.[33]

Raman spectroscopy was carried out on individual $CrCl_3$ flakes using a commercial confocal microscope (Horiba Xplora Plus) with an excitation wavelength of 532 nm. Photoluminescence spectroscopy in spectral and time-resolved mode was carried out at room temperature using a commercial flash-photolysis laser system (Edinburgh Instruments LP920) with excitation wavelengths 532 nm and 700 nm and pulse width of ~ 7 ns.

## 3. Results and Discussion

### 3a. Electronic Spectroscopy of $CrCl_3$

Figure 1c-1d shows the characterization of our $CrCl_3$ samples using Raman, steady-state DRS, and pulsed PL spectroscopy. Figure 1b is a microscope image of a typical $CrCl_3$ flake with relatively large flat regions. The Raman spectra measured on this sample and in Figure 1c show the expected fingerprint vibrational modes based on prior reports for this material.[34] We infer from this comparison that our samples are representative of bulk $CrCl_3$ in terms of crystal structure and chemical composition.

Further characterization of the samples comes from the static absorption and pulsed PL spectra shown in Figure 1d. The DRS spectra exhibit two broad LFE peaks at 1.75 eV and 2.5 eV, which have been previously assigned as transitions from the $^4A_2$ ground state into the $^4T_2$ and $^4T_1$ excited states.[14,15] The transition energies and line shapes agree well with several previous reports that assign the substantial broadening of the LFE peaks due to trigonal distortions of the octahedral environment around the $Cr^{3+}$ cation.[14,18] The pulsed PL spectrum shows a peak at 1.4 eV that corresponds well to prior studies of $CrCl_3$.[6]

From Figure 1d, we can directly assess the significant electron-phonon coupling in $CrCl_3$ by noting the Stokes shift between the lowest LFE peak in absorption and the PL emission with 532 nm excitation. The Stokes shift of ~0.35 eV indicates that substantial vibrational relaxation occurs immediately after the creation of the LFE. This shift is a common feature of LFE's in $Cr^{3+}$ compounds, and it is important to consider the substantial role of electron-phonon coupling when assessing the general many-body relaxation in these materials. The Stokes shift for $CrCl_3$ is slightly smaller than the 0.43 eV reported for $CrI_3$.[20] It is possible that these slight differences could be accommodated within a simple Frohlich polaron model. However it is noteworthy that Frohlich polarons in hybrid lead halide perovskite materials show the opposite trend to what we see for $CrX_3$ in electron-phonon coupling in going from iodide to chloride.[35] The quantitative details of polaronic effects in the chromium trihalides are an important topic for future focus, and indeed, there is evidence that LFE's in $CrI_3$ should be regarded as "exciton-polarons".[36]

The PL spectrum for $CrCl_3$ is identical for excitation wavelengths of either 700 nm or 532 nm as shown in Figures 2a and 2b. We have not observed any emission from the higher energy $^4T_1$ LFE state, presumably because it decays very rapidly into the lower $^4T_2$ energy state. Luminescence can result from either the $^4T_2$ or $^2E$ excited state depending on ligand field strength modulation (*Dq/B*) as described by the $d^3$ configuration Tanabe-Sugano diagram. Owing to the peak broadness of the observed emission feature (~70 nm)[17] and weak-field *Dq/B* value as evaluated from the steady-state absorption spectrum ($Dq/B = 2.26$), we assign this luminescence as a spin-allowed $^4T_2 \rightarrow {}^4A_2$ emission. The linewidth of the emission spectrum is expected to be dominated by the vibronic relaxation in the excited state using the textbook quantitative analysis[23] that was also recently shown to apply emission spectra for $CrI_3$ [20].

In Figure 2c we show the incident laser power dependence of the integrated PL intensity. A quantitative fit to the data in Figure 2a indicates that the energy-integrated PL intensity scales approximately as the square root of power for 532 nm excitation. This relatively slow (sublinear) growth of the PL intensity with increasing laser pulse energy, is an indication of the complex processes involved in the excited state population that could be fit to a number of different models, including piece-wise linear fits. However, in anticipation of our ultimate interpretation, this observation is common in systems exhibiting EEA [37,38] since the EEA process removes increasingly more emitters as their density increases with incident laser power. Beyond the supporting evidence in Figure 2c, We present definitive evidence that EEA is the dominant relaxation process by considering TRPL in the next section.

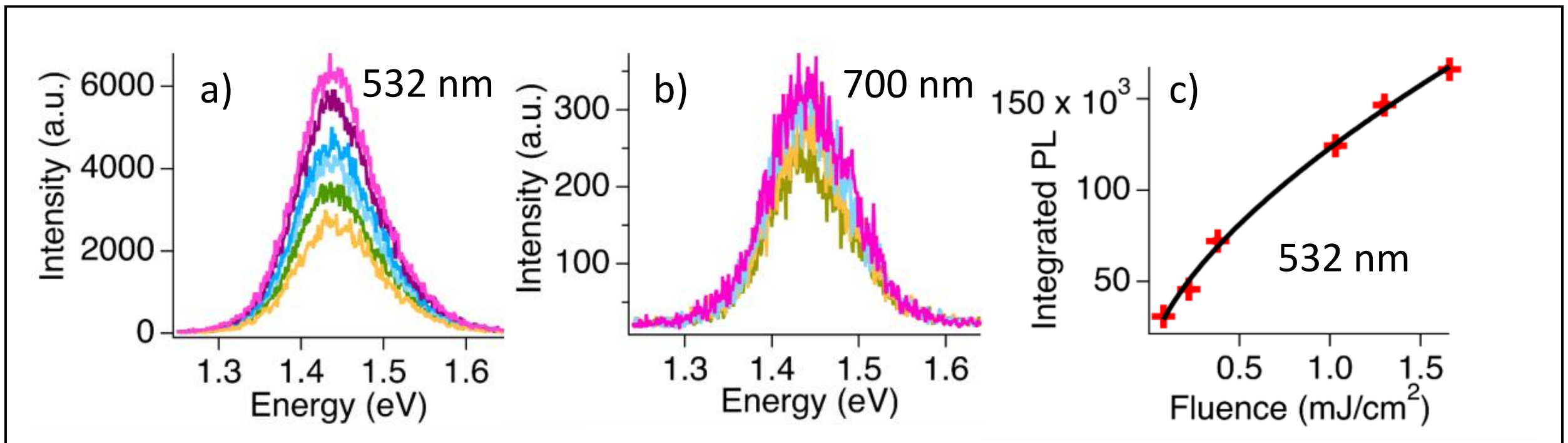

**Figure 2.** a) PL spectrum with 532 nm excitation and; b) with 700 nm excitation; c) The incident Fluence dependence of the integrated PL spectrum from part a (crosses) showing growth with a best fit power law (approximately square root scaling) as the black solid line.

### 3b. Time Resolved Photoluminesence

In pulsed laser experiments on $CrCl_3$, the PL intensity decays over the course of several microseconds, as shown in Figure 3. In Figure 3a, the population relaxation at 532 nm excitation wavelength can be seen. Similar time-dependent behavior can also be seen for excitation at 700 nm as described below. For tr-PL data at room temperature on $CrCl_3$, such as shown in Figure 3a, it is impossible to fit the decaying PL intensity with a single exponential function as expected for spontaneous decay of the excited state by photon or phonon emission or interaction with defects. This means that the mechanism of LFE population decay at 300K is not first-order exciton recombination, in contrast with the recent observation of simple exponential decay in tr-PL in this material at 4K.[32]

There are many possibilities for understanding complex relaxation kinetics in tr-PL experiments. An important insight in our data comes from an inspection of the dependence of decay kinetics on incident laser power. In Figure 3b we see that the initial PL intensity grows sublinearly with increasing incident laser Fluence. We would expect the initial intensity to grow linearly with light fluence until saturation for a linear process. The slow growth suggests that there are direct interactions between LFE's that deplete the initial population at high densities. These interactions will be shown below to be direct EEA.

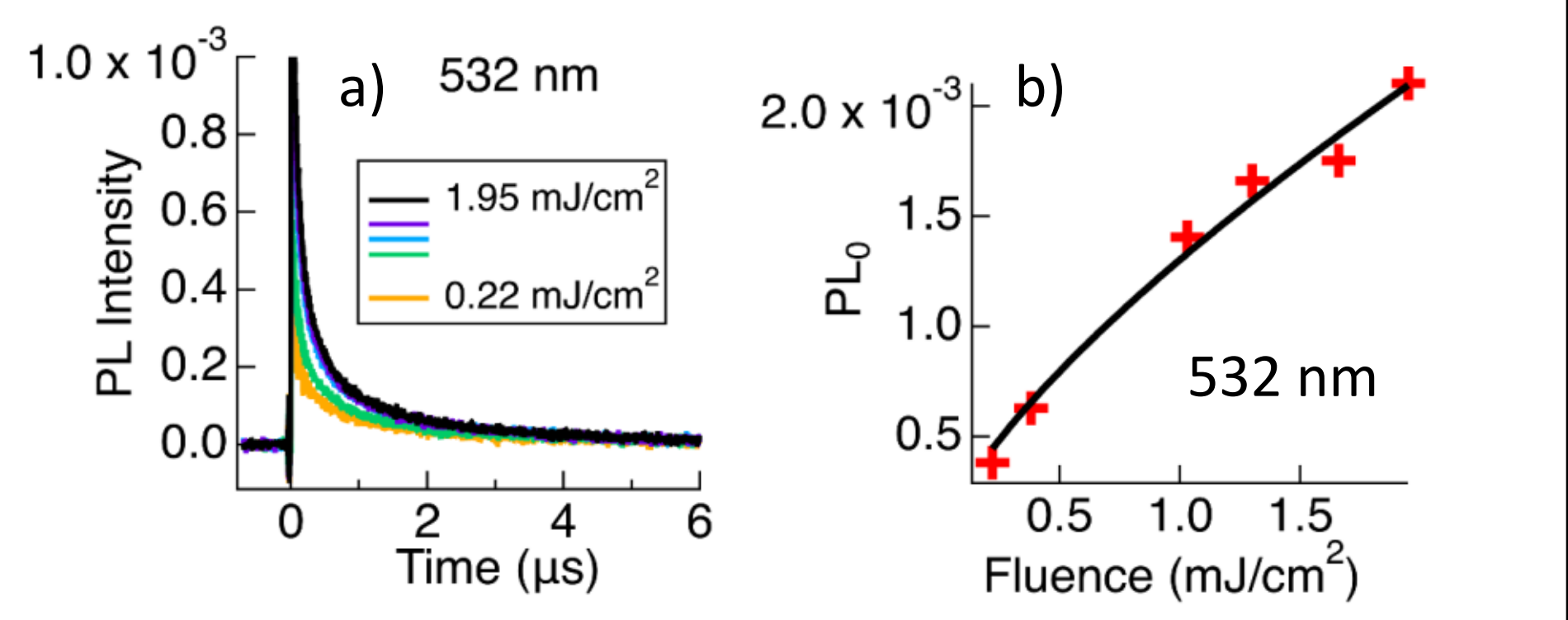

**Figure 3.** a) Time-resolved PL decay traces measured at 1.45 eV (532 nm excitation) at different incident laser fluences; b) Fluence-dependence of the initial PL intensity from the traces in part a) showing sublinear growth.

To address the EEA hypothesis more quantitatively, we plot the decay data using a standard method for analyzing pure second-order kinetics, Figure 4. Here, the linear scaling of inverse PL intensity with time (for early times) agrees with expectations for the pure second-order kinetic equation describing EEA. In general, population decay kinetics, including both first-order decay and second-order EEA effects, can be described by the following differential equation for exciton population *N(t)*:

$$\frac{dN}{dt} = -kN - \gamma N^2 \qquad (1)$$

Where $k$ is the spontaneous exciton decay rate that includes both radiative and nonradiative processes and $\gamma$ is the EEA rate constant. Equation 1 has an analytical solution that has been used to fit tr-PL data in materials like $WS_2$[39], $CrI_3$,[36] and 2D lead halide perovskites[40] and extract numerical values for both $k$ and $\gamma$. Our fits to the general analytical solution to Eq. (1) yield spontaneous decay lifetimes that are sometimes longer than our experimental observation time and are also very sensitive to the precise range of data chosen to be fit. We infer from this that the spontaneous decay lifetime for LFE's in bulk $CrCl_3$ is quite long (10-100 μs; similar to the 4K value of 7.4 μs [32]) and thus the first term in equation 1 is not relevant to quantifying our tr-PL observations.

Instead, we fit the variant of Eq. (1) that only includes the second EEA term, as shown in Figures 4a and 4b for 532 nm and 700 nm excitation, respectively. The exact solution for this situation is given by

$$N(t) = \frac{N_0}{1+\gamma N_0 t}, \qquad (2)$$

where $N_0$ is the initial LFE population. The physical meaning of this analysis is that the EEA processes *dominate* the LFE population decay in bulk $CrCl_3$. Other processes, such as spontaneous or defect-mediated radiative or non-radiative decay, are evidently too unlikely to be relevant on the timescales considered here. In early work on single-layer $MoS_2$, it was noted that EEA dominated the population relaxation on ultrafast timescales of tens of picoseconds.[41] Our measurements are on much longer timescales but access a similar qualitative scenario where population decay does not occur to any significant extent spontaneously but only when LFE's encounter one another and annihilate.

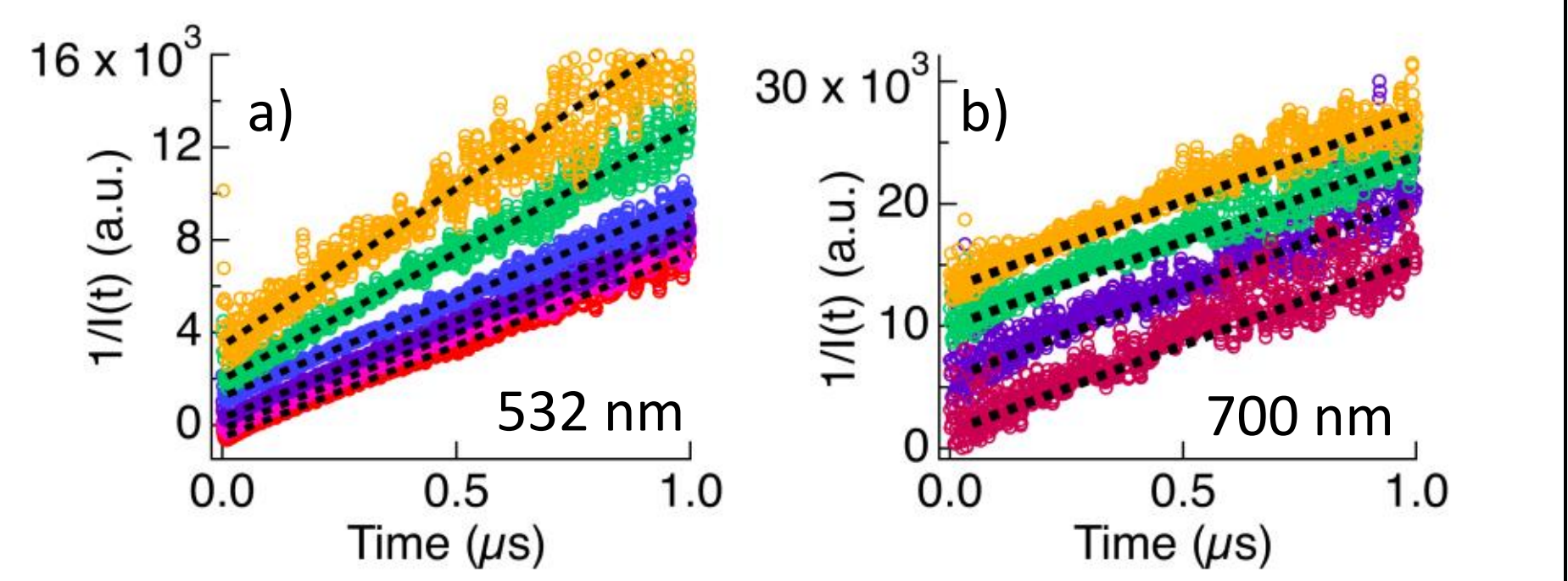

**Figure 4.** a) Plot of Inverse PL intensity at 1.45 eV (532 nm excitation) as a function of time showing linear growth indicative of pure second order kinetics. The fluences *increase* from top curve to bottom curve in the same range as in Figure 3a) (0.22, 0.38, 1.03, 1.30, 1.66, 1.95 mJ/cm$^2$); b) Inverse PL intensity versus time at 1.45 eV (700 nm excitation) for a range of fluences increasing from bottom to top (0.03, 0.05,0 .15, 0.30 mJ/cm$^2$).

### 3c. Annihilation of Ligand Field Excitons

The EEA process in solids can be depicted schematically as shown in Figure 5a. It is often described as "Auger-like" in the sense that it involves at least 3 different states: the ground state, the lowest excited state, and a higher-lying excited state.[42] Two excitons, with electrons in the lowest excited state and holes in the ground state, exchange energy when one decays and transiently excites the electron on the second exciton into the higher-lying excited state. It rapidly decays back, leaving only one exciton, possibly with excess kinetic energy.[43]

The coupling between two excitons generally relates to the square of the transfer integral between excitons that describes energy exchange. Recently it has been argued that this generates a simple and "universal" connection between spontaneous decay lifetimes and EEA rate constants where the EEA rate is inversely proportional to the intrinsic radiative decay lifetime.[44] Aggregated experimental data show statistical correlations consistent with the trend expected from this simple universal formula. Here the word "universal" should be understood similarly to the "universal" curve describing the inelastic mean free path of electrons in solids as a function of their kinetic energy. Namely, as a way to assess trends across numerous diverse materials even though the details for each individual material are likely to lead to substantial fluctuations away from the trend.

To place our measurements in the context of this perspective on EEA, we consider the data aggregated by Uddin et al.[44] for 3D materials. Figure 5b shows this data in blue circles on a double logarithmic plot of EEA rate constant versus exciton lifetime that illustrates the general trend of decreasing EEA rate with lifetime even though the fluctuations between different materials with similar time constants are very large. For bulk $CrCl_3$, quantifying the EEA rate from tr-PL in physical units requires calibration of the PL signal in terms of initial exciton density. In particular, it requires knowledge of the PL quantum yield (PLQY) for this material, which is currently not known. However, the large spread in the aggregated literature data in Figure 5b suggests the value of seeking order of magnitude estimates for the EEA rate in bulk $CrCl_3$.

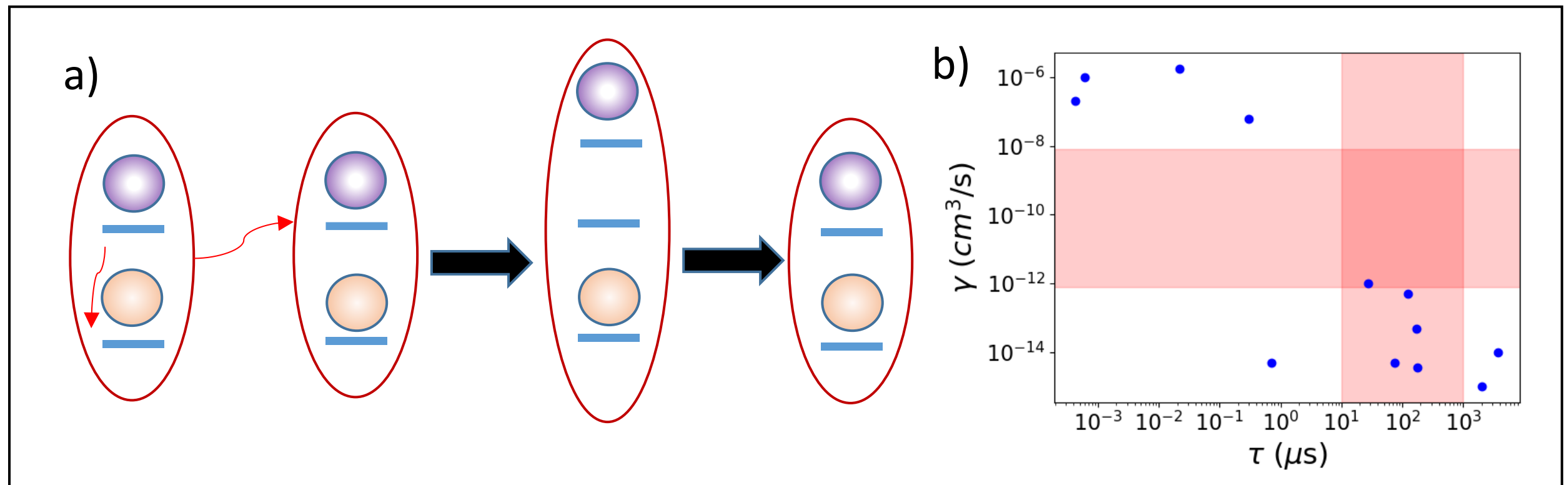

**Figure 5.** a) Schematic depiction of the EEA process where two excitons exchange energy via internal conversion to make a more highly excited intermediate state. This state can eventually relax back to a single exciton; b) Blue circles plot the survey of 3D material EEA rates ($\gamma$) and lifetimes ($\tau$) originally tabulated in Uddin et al.[44] The pink shaded regions indicate the plausible values ascertained from our tr-PL study of bulk $CrCl_3$. The intersection of the two regions suggests a possible enhancement of EEA rate compared to universal trend expectations.

For both LFE's in $CrCl_3$, the absorption coefficient has been quantified[14] as about 2000 $cm^{-1}$. For an incident laser energy of 1 mJ/pulse, at 2.33 eV photon energy, this results in an estimated $2.7 \times 10^{15}$ photons/pulse. Assuming that all photons are absorbed within a depth of one inverse of the absorption coefficient over an area of about $1 cm^2$, the volume density of photons is $5 \times 10^{18}$ $cm^{-3}$. If each of these photons is assumed to generate an LFE, we can calibrate the PL intensity to extract a range of possible EEA rate constants from the slope in Figure 4 for different PLQY values from 1 down to $10^{-4}$. The intersection of pink-shaded regions in Figure 5b shows likely regions of $\gamma$ and $\tau$ for bulk $CrCl_3$. In particular, we note that the values for $\gamma$ are likely to be in the range of $8 \times 10^{-13}$ $cm^3/s$ up to $8 \times 10^{-9}$ $cm^3/s$, depending on PLQY. The larger calculated EEA rates correspond to the smaller assumed PLQY values, which we consider more likely.

Given the large variability between different materials, the possible EEA rates are somewhat consistent with the universal trend.[44] But there is some indication of an upward deviation from the trend, particular given the likelihood of a PLQY substantially less than unity. Enhanced EEA rates compared to expectations in the universal model could arise from the effects of excess kinetic energy on exciton diffusion. At high exciton generation rates, the effective diffusion coefficient can be enhanced due to the creation of "hot" excitons through the EEA process that can then diffuse more rapidly.[43] This gives an apparently larger EEA rate than expected from the universal model based on fully thermalized materials properties.

Indeed, regardless of the detailed quantification of EEA rates, it is clear that excitons in bulk $CrCl_3$ must diffuse readily to establish the dominant EEA interactions. Such considerations have been made in describing very long-lived exciton populations in $TbPO_4$[45] and $GdCl_3$[46] which have EEA rate constants of $5 \times 10^{-15}$ $cm^3/s$ and $4.7 \times 10^{-14}$ $cm^3/s$ respectively (these values are included in the blue data in Figure 5b). In fact, these two materials are perhaps the closest literature connection to $CrCl_3$ for which excited state relaxation has been quantified in that they also involve highly localized excitons on single metal cation sites. In the case of the rare earth cations, the LFE arises from excitations between ligand field-split $f$ orbitals instead of the $d$ orbitals in $CrCl_3$ but overall the situation is quite similar. In both cases, the LFE's are inferred to move by incoherent hopping with a 1D diffusion constant related to the EEA rate by $\gamma = 8\pi D\langle R\rangle$, where $\langle R\rangle$ is the average radius at which EEA interactions occurs. In prior work, the value for $\langle R\rangle$ was taken to be comparable to the cation spacing in the solid (~0.5 nm).[45,46] We expect the same to be true for $CrCl_3$ but we can also take advantage of recent theoretical calculations of exciton wave functions to directly estimate the size of the LFE, which was reported to be essentially confined within a unit cell.[21] Within a layer, the nearest-neighbor Cr spacing is 0.35 nm, and the lowest energy spin-allowed LFE is represented as essentially spanning this distance. Thus, the diffusion coefficient can be estimated to be in the range of $9x10^{-7}$ $cm^2/s$ to $9x10^{-3}$ $cm^2/s$ which is larger than the two related rare earth materials.

An additional effect in our data is the observation of a slight decrease in EEA rate constant with incident excitation density at 532 nm excitation, as is apparent by inspection of Figure 4a. This dependence can be understood as the result of laser-induced heating of the $CrCl_3$ sample at the higher incident fluences. The EEA rate is known to vary inversely with temperature due to coupling of the exciton population to the bosonic bath of phonons.[47,48] Lower temperatures lead to larger EEA rates due to the higher population of excitons with smaller center of mass momenta under these conditions. This is a large-scale macroscopic heating effect distinct from the possible transient hot exciton diffusion effects described earlier.[43]

Relatedly, our tr-PL observations at room temperature in Figure 4 contrast sharply with the single exponential tr-PL decay at low temperatures [32]. Since EEA rates typically increase with increasing temperature,[48] it is not surprising that the impact of LFE annihilation is more notable in high-temperature experiments. However, in our observation, EEA is the dominant contribution to excited state relaxation. This points to a relatively temperature-independent first-order LFE lifetime that is evidently long enough to allow the EEA rate to become relatively more significant at elevated temperatures.

Time-resolved photoluminescence of exfoliated flakes of $CrI_3$ also shows evidence of EEA effects.[36] However, in that system, the first-order lifetime is far shorter (~2 ns) than the recent tr-PL study on bulk $CrI_3$ crystals that indicates lifetimes closer to 1 μs.[32] In addition, the EEA rates for the flakes are estimated from global fits to the solution of Eq. (1) to be very small. We hypothesize that the contrast between this related system and the bulk $CrCl_3$ considered here arises from the strong influence of the supporting substrate on excited state populations in ultrathin flakes. The substrate can serve as a source of "defect-induced" first-order recombination that reduces the excited state population lifetime before EEA effects can become dominant. In the bulk crystals we study, LFE's live long enough that the only source of population relaxation is EEA.

## 4.Summary and Conclusions

In summary, we have identified exciton-exciton annihilation as the dominant relaxation mechanism for ligand field exciton populations at room temperature in bulk crystals of $CrCl_3$. The intrinsic lifetime of these excitons is too long to quantify in our experiments directly, but is likely to be in the range of 10-100 μs. The EEA rate constant could be crudely estimated from our experiments and may be in the range expected from the universal trend relating it to spontaneous decay lifetime. However, our data are also consistent with a substantially larger EEA rate than predicted by this trend, and refinement of the measurement of EEA rates is an important future goal.

The dominance of the EEA process in controlling excited state populations in $CrCl_3$ has critical implications for optoelectronic control. For example, EEA effectively prohibits access to very high excitation density. Future work could mitigate this effect by using a dielectric environment or strain. However, it is also worth noting that, even in the presence of dominant EEA, substantial excited state populations in $CrCl_3$ persist for a time scale of order 1 μs. This may be long enough to allow manipulations of these excited states in novel scenarios like photoconductive devices in van der Waals heterostructures. From this perspective, the LFEs in $CrCl_3$ point to the chromium trihalides as a valuable optoelectronics materials class.

**Acknowledgment.** ATB and FNC were supported by the U.S. Department of Energy, Office of Science, Office of Basic Energy Sciences, Award Number DE-SC0011979.

## 5.References